\documentclass[conference]{IEEEtran}
\usepackage[utf8]{inputenc}
\usepackage{times,graphics, dsfont, epsfig,amsmath,xspace,endnotes,pifont,multirow,rotating,listings,amssymb,algorithmic,color,caption,nicefrac,adjustbox,tabularx,mathtools, algorithmic,algorithm}
\usepackage{graphicx}
\usepackage{tabularx}

\newcolumntype{L}[1]{>{\raggedright\arraybackslash}p{#1}} 
\newcolumntype{C}[1]{>{\centering\arraybackslash}p{#1}} 
\newcolumntype{R}[1]{>{\raggedleft\arraybackslash}p{#1}} %
\usepackage{steinmetz}
\usepackage{color}
\usepackage[dvipsnames]{xcolor}
\usepackage{amsmath}
\usepackage{amssymb}
\usepackage{float}
\ifCLASSOPTIONcompsoc
    \usepackage[caption=false, font=normalsize, labelfont=sf, textfont=sf]{subfig}
\else
\usepackage[caption=false, font=footnotesize]{subfig}
\fi
\usepackage{graphicx}
\usepackage[letterpaper, left=0.625in, right=0.625in, bottom=1in, top=0.75in]{geometry}

\usepackage{tikz}
\usepackage{amsthm}
\newtheorem{thm}{Theorem}

\newtheorem{defn}[thm]{Definition}

\usepackage{steinmetz}
\usepackage{tikz}

\newcommand{\s} {{\bf{s}}}

\newcommand{\nariman}[1]{\textcolor{red}{#1}}
\newcommand{\amir}[1]{\textcolor{blue}{#1}}

\author{
\IEEEauthorblockN{Nariman Torkzaban}
\IEEEauthorblockA{\textit{University of Maryland, College Park}\\
College Park, MD\\
narimant@umd.edu}
\and
\IEEEauthorblockN{Mohammad A. (Amir) Khojastepour}
\IEEEauthorblockA{\textit{NEC Laboratories,
America}\\
Princeton, NJ\\
amir@nec-labs.com}
\and
\IEEEauthorblockN{John S. Baras}
\IEEEauthorblockA{\textit{University of Maryland, College Park}\\
College Park, MD\\
baras@umd.edu}
}

\def\BibTeX{{\rm B\kern-.05em{\sc i\kern-.025em b}\kern-.08em
    T\kern-.1667em\lower.7ex\hbox{E}\kern-.125emX}}

\newcommand\copyrighttext{%
  \footnotesize \textcopyright 2022 IEEE. Personal use of this material is permitted.
  Permission from IEEE must be obtained for all other uses, in any current or future
  media, including reprinting/republishing this material for advertising or promotional
  purposes, creating new collective works, for resale or redistribution to servers or
  lists, or reuse of any copyrighted component of this work in other works.
  }
\newcommand\copyrightnotice{%
\begin{tikzpicture}[remember picture,overlay]
\node[anchor=south,yshift=10pt] at (current page.south) {\fbox{\parbox{\dimexpr\textwidth-\fboxsep-\fboxrule\relax}{\copyrighttext}}};
\end{tikzpicture}%
}
\begin{document}

\title{Multi-user Beam Alignment in Presence of Multi-path}



\maketitle
\copyrightnotice
\begin{abstract}
To overcome the high path-loss and the intense shadowing in millimeter-wave (mmWave) communications, effective beamforming schemes are required which incorporate narrow beams with high beamforming gains. The mmWave channel consists of a few spatial clusters each associated with an angle of departure (AoD). The narrow beams must be aligned with the channel AoDs to increase the beamforming gain. This is achieved through a procedure called \emph{beam alignment (BA)}. Most of the BA schemes in the literature consider channels with a single dominant path while in practice the channel has a few resolvable paths with different AoDs, hence, such BA schemes may not work correctly in the presence of multi-path or at the least do not exploit such multipath to achieve diversity or increase robustness.  
In this paper, we propose an efficient BA schemes in presence of multi-path. The proposed BA scheme transmits probing packets using a set of scanning beams and receives the feedback for all the scanning beams at the end of probing phase from each user. We formulate the BA scheme as minimizing the expected value of the average transmission beamwidth under different policies. The policy is defined as a function from the set of received feedback to the set of transmission beams (TB). In order to maximize the number of possible feedback sequences, we prove that the set of scanning beams (SB) has an special form, namely, \emph{Tulip Design}. Consequently, we rewrite the minimization problem with a set of linear constraints and reduced number of variables which is solved by using an efficient greedy algorithm.

\end{abstract}


%
\IEEEpeerreviewmaketitle

\section{Introduction}
In pursuance of larger bandwidth that is required for realizing one of the main promises of 5G, i.e. enhanced mobile broadband (eMBB), millimeter wave (mmWave) communications is a key technology due to abundance of unused spectrum available at mmWave frequency ranges \cite{Bus18}. However, high path loss and poor scattering associated with mmWave communications leads to intense shadowing and severe blockage, especially in dense urban environments. These are among the major obstacles to increase data rate in such high frequency bands. To tackle these issues effective beamforming (BF) techniques are required to avoid the power leakage to undesired directions using directional transmission patterns, i.e., narrow beams \cite{Kut16}. Furthermore, several experimental results demonstrate that the mmWave channel usually consists of a few components (a.k.a spatial clusters) \cite{Akd14}. Therefore, it is essential to align the devised narrow transmission beams with the direction of the channel components. The problem of aligning the directions of the beams with the angle of departure (AoD) associated with clusters of the channel, is termed as the \emph{beam alignment (BA)} problem. In the literature the beam alignment problem is also indexed as \emph{beam training} or \emph{beam search}. Devising effective beam alignment schemes is essential since a slight deviation of the transmitted beam AoDs from the mmWave channel clusters may result in a severe drop in the beamforming gain \cite{Nit15}\cite{Shah19}. 

Beam alignment schemes may be categorized as \emph{exhaustive search (ES)} and \emph{hierarchical search (HS)}. Under the ES scheme, a.k.a beam sweeping, the angular search space is divided into multiple angular coverage intervals (ACIs) each covered by a beam. Then the beam with the highest received signal strength at the receiver is chosen \cite{Bar16}\cite{Gio16}. 
To yield narrower beams in ES, the number of beams increases which results in larger beam sweeping overhead.
The HS scheme lowers the overhead of the beam search by first scanning the angular search space by coarser beams and then gradually finer beams \cite{Des14}\cite{Hus17}\cite{Noh17}. The BA procedure may happen in one of the two modes, i.e. \emph{interactive} BA (I-BA), and \emph{non-interactive} BA (NI-BA). In the NI-BA mode, the transmitter sends the scanning packets in the scanning phase and receives the feedback from the users after the scanning phase is over, while in the I-BA mode the transmitter receives the feedback for the previously transmitted scanning pilots during the scanning phase and can utilize this information in the rest of the scanning phase. 
Most of the prior art on I-BA are limited to single-user scenarios while NI-BA schemes can handle multi-user scenarios as the set of scanning beam does not change or depend on the received feedback from the users.

In this paper, we address the problem of beam alignment in multipath environment. We formulate the BA scheme as minimizing the expected average transmission beamwidth under different policies. The policy is defined as a function from the set of received feedback to the set of transmission beams (TB). In order to maximize the number of possible feedback sequences, we prove that the set of scanning beams (SB) has an special form, namely, \emph{Tulip Design}. Consequently, we rewrite the minimization problem with a set of linear constraints and reduced number of variables which is solved by using an efficient greedy algorithm.

\textbf{Notations.} Throughout this paper, "$\setminus$" denotes the set minus operation, $[b]$ denotes the set of all integers greater than or equal to zero and less than $b$. $\oplus$, and $\ominus$ denote the summation and subtraction operations in the base $b$. $|.|$ may denote the cardinality if applied to a set or the Lebesgue measure if applied to an interval.

The remainder of the paper is organized as follows. Section~\ref{sec:desc} describes the system model. In Section~\ref{sec:problem} we formulate the BA problem formulation and propose our solutions in section~\ref{sec:proposed}. Section~\ref{sec:evaluation} presents our evaluation results, and finally, in Section~\ref{sec:conclusions}, we highlight our conclusions.

\section{System model} 
\label{sec:desc}
We consider a mmWave communications scenario with a single base station (BS) and an arbitrary number of mobile users (MUs), say $N$, where prior knowledge on the value of $N$ may or may not be available at the BS. The BA procedure aims at obtaining the accurate AoDs corresponding to the downlink mmWave channel from the BS to the users. Under the BA procedure, the BS transmits probing packets in different directions via various \emph{scanning beams (SBs)} and receives feedback from all the users, based on which the BS computes a \emph{transmission beam (TB)} for each user. 

\subsection{Channel Model}
Unlike prior art, we consider multipath in the transmission from the BS to the MUs. More precisely, we assume the mmWave channel from the BS to each MU contains a maximum of $p$ resolvable paths where each \emph{resolvable path} corresponds to a possible AoD of the channel. Let $\Psi_{j}= \left\{\Psi_{i j}\right\}_{i=1}^p$ denote the random AoD vector corresponding to the channel between the BS and the $j^{th}$ MU, where $\Psi_{i j}$ represents the AoD of the $i^{th}$ path. Denote by $f_{\Psi_{j}}\left(\psi_{1 j}, \ldots, \psi_{p j}\right)$, defined over $\mathcal{D} \subset(0,2 \pi]^{p}$, the probability density function (PDF) of $\Psi_j$. The PDF $f_{\Psi_{j}}(.)$ encapsulates the knowledge about the AoD of the $j^{th}$ user prior to the BA procedure, or may act as a priority function over the angular search domain. Such information may be inferred from previous beam tracking, training, or alignment trials. A uniform distribution is tantamount to the lack of any prior knowledge or priority over the search domain. 

\subsection{Beamforming Model}
We consider a multi-antenna base station with an antenna array of large size realizing beams of high resolution. For power efficiency, we assuming hybrid beamforming techniques are in effect in the BS deploying only a few RF chains. Further, we adopt a \emph{sectored antenna model} where each beam is modeled by the constant gain of its main lobe, and the angular coverage interval (ACI) it covers. Such models are widely adopted in the literature for modeling the beamforming gain and the directivity of mmWave transmitters. 

\subsection{Time-slotted System Model}
We consider a system operating under the time division duplex (TDD) and the NI-BA schemes, with frames of length $T$. Each frame consists of $T$ equal slots. In each frame, the first $b$ slots are dedicated to the transmission of the probing packets, denoted by \emph{scanning time-slots (STS)} and the next $d$ slots denoted by \emph{feedback time-slots (FTS)} are allocated to receiving the users feedback that may arrive through a side channel or according to any random access mode. Finally, the last $T+b+d$ slots are reserved for data transmission, namely \emph{data transmission time-slots (DTS)}.
\begin{figure}
    \centering
    \includegraphics[width=0.9\linewidth]{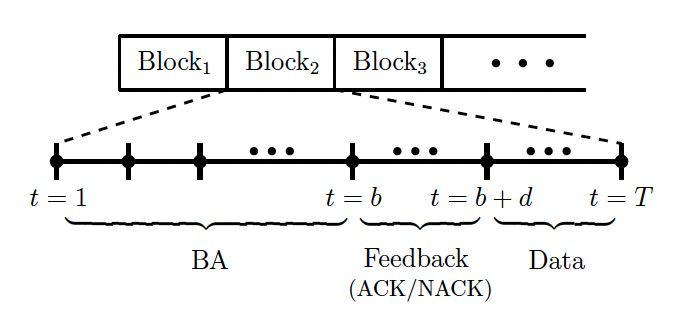}
    \caption{Time-slotted System Model}
    \label{fig:system}
\end{figure}

\subsection{Beam Alignment Model}
The objective of the BA scheme is to generate narrowest possible TBs for the data transmission phase for each user to produce beams of higher gain and quality. In other words, utilizing the feedbacks provided by the users in the FTS to the SBs transmitted by the BS in the STS, the BS aims at localizing the AoD of each user to minimize the \emph{uncertainty region (UR)} for each AoD. Let $\mathcal{B}=\left\{\Phi_{i}\right\}_{i=1}^{b}$ be the set of STS scanning beams where $\Phi_{i}$ denotes the ACI of the SB sent over time-slot $i \in [b]$. The feedback provided by each user to each SB is binary. If the AoD corresponding to at least one of the resolvable paths in the channel from the BS to the MU is within the ACI of the SB, then the MU will receive the probing packet sent via that SB and feeds back an acknowledgment (ACK). Otherwise, the feedback of the MU will be considered as a negative acknowledgment (NACK) indicating none of the user AoD's lie in the ACI of the SB. Once the FTS ends, the BS will determine the TBs using the SBs and the feedback sequences provided by the users according to the BA \emph{policy}. The BA policy is formally defined as a function from the set of feedback sequences to the set of TBs. 

In the next section, we first elaborate on the BA policy and then provide the BA problem formulation. 
\section{Problem Formulation}
\label{sec:problem}

\subsection{Preliminaries}

The BA policy determines how the direction of the TBs is computed. This decision naturally considers the UR of the AoDs of each user channel. In this paper, we consider four different policies that differ based on how they define the URs of the AoDs and whether they require the exact number of spatial clusters or not.  

\subsubsection{general policies}
We define two general policies that do not require any information regarding the number of spatial clusters, namely i) spatial diversity (SD) policy, ii) beamforming (BF) policy. The SD policy aims at generating TBs with minimal angular span that cover all the angular intervals that may contain a resolvable path, while the promise of the BF policy is to generate TBs that cover at least one resolvable path but further reduce the angular span of the TBs. The advantage of the SD policy to the BF policy is its resilience against the potential failure or blockage of one or some of the spatial clusters as long as at least one resolvable path remains, while the BF policy has the advantage of producing much higher beamforming gains compared to that of the SD policy but it is vulnerable to path blockage. This will introduce an interesting trade-off between connectivity maintenance and high beamforming gain. 

\subsubsection{path-based policies}
If the exact number of spatial clusters, $p$, is known, each of the above policies may be improved by further lowering the span of the resultant TBs. We denote the corresponding two new policies by $p$-SD policy and $p$-BF policy respectively. 

In the following, we state the expressions for the URs corresponding to each of the mentioned SD and BF policies. Let $B^j_{\mathcal{P}}(\mathcal{B}, \mathbf{s})$ denote the UR of the $j^{th}$ user providing the feedback sequence $\s$ under the policy $\mathcal{P} \in \{\text{SD}, \text{BF}, $p$-\text{SD},  $p$-\text{BF}\}$  and the SB set $\mathcal{B}$. For instance, $B^j_{SD}(\mathcal{B}, \mathbf{s})$ is the minimal angular span that covers all the resolvable paths for the $j^{th}$ user. Similarly, $B^j_{BF}(\mathcal{B}, \mathbf{s})$ is the minimal angular span that covers at least  one resolvable path for the $j^{th}$ user. Further, let the \emph{positivity set} $A^j(\s) \subseteq [b]$ be the set of all indices corresponding to the SBs that are acknowledged by the $j^{th}$ user. Define the \emph{negativity set} $N^j(\s) \subseteq [b]$ in a similar fashion for the not acknowledged SBs. 
To facilitate the statement of the URs and the subsequent following discussions we define the notion of the \emph{component beam (CB)}. The CB $\omega_A$ is defined as, 
\begin{align}
    \omega_A = \cap_{i \in A} \Phi_{i} \backslash \cup_{A \subset T \subset[b], A \neq T} \cap_{i \in T} \Phi_{i}
\end{align}

It is straightforward to show that $\omega_{A} \cap \omega_{T}=\varnothing$ for any $A\neq T$, and $\Phi_{i}=\cup_{A, i \in A} \omega_{A}$ for all $i \in [b]$. We define the CB set as $\mathcal{C}=\left\{\omega_{A}, \omega_{A} \neq \varnothing, A \subset[b]\right\}$. Obviously, $\mathcal{B}$ can be generated form $\mathcal{C}$ and vice versa.

Note that if the $j^{th}$ user sends an ACK in response to the SB $\Phi_i$, this would mean that $\Theta_{i}(\mathbf{s})\doteq \Phi_{i}$ has at least one resolvable path. On the other hand, a NACK would mean that no resolvable paths reside in $\Phi_{i}$ and therefore,  any resolvable path should exist in $\Theta_{i}(\mathbf{s})\doteq \mathcal{D}-\Phi_{i}$, and $B^j_{\mathcal{P}}(\mathcal{B}, \mathbf{s}) \in \mathcal{D}-\Phi_{i}$ for the above-mentioned policies. Having this in mind, we can explicitly express the uncertainty region for the general policies as follows.
\begin{align}
    &B^j_{\text{SD}}(\mathbf{s})=\left(\cup_{i \in A(\mathbf{s})} \Theta_{i}(\mathbf{s})\right) \cap\left(\cap_{i \in N(\mathbf{s})} \Theta_{i}(\mathbf{s})\right)\\
    &B^j_{\text{BF}}(\mathbf{s})=\Theta_{k}(\mathbf{s}) \cap\left(\cap_{i \in {N}(\mathbf{s})} \Theta_{i}(\mathbf{s})\right)
\end{align}
where $k={\arg \min }_{{l \in {A}(\mathbf{s})}} \left |\Theta_{l}(\mathbf{s}) \cap\left(\cap_{i \in N(\mathbf{s})} \Theta_{i}(\mathbf{s})\right) \right|$. For the path-based policies, having the luxury of the knowledge on the exact value of $p$, we can improve the SD and the BF policies to $p$-SD and $p$-BF, respectively. For the simplicity of presentation, we only express the improved policies for $p=2$.

We define $\mathcal{W}_A = \{ \{C, C'\}  \text{ s.t. }  \omega_C, \omega_{C'} \in \mathcal{C}, \text{ } C \cup C' = A \}$, $\mathcal{V}_A = \cup_{V \in \mathcal{W}_A} V $and $n = |\mathcal{W}_A|$. Let $\bigotimes \mathcal{W}_A$ denote the Cartesian product of all elements of $\mathcal{W}_A$ where each element of $\bigotimes \mathcal{W}_A$ is a n-tuple. For a n-tuple $T = (t_1, t_2, \ldots, t_n)$, we define $\textsc{union}(T) = \cup_{i=1}^n t_i$. We have
\begin{align}
    &B^j_{2-\text{SD}}(\mathbf{s})= \bigcup_{V \in \mathcal{V}_A} \omega_V\\
    &B^j_{2-\text{BF}}(\mathbf{s})= \textsc{union}(T^*), \text{}
    T^* = \arg\min_{T \in \bigotimes \mathcal{W}_A} |\textsc{union}(T)|
\end{align}
Note that for the special case of $p=1$ all the mentioned policies collapse into one. Next, we will present the BA problem formulation. 

\subsection{Problem Formulation} 
We assume there are $N$ users that are prioritized according to the weight vector $\left\{c_{j} \geq 0\right\}_{j=1}^{N}, \sum_{j=1}^{N} c_{j}=1$. Let $\mathcal{U}=\left\{u_{k}\right\}_{k=1}^{M}$ denote the range of the policy function $B^j_{\mathcal{P}}(\mathcal{B}, \mathbf{s})$. In other words, the TBs resulting from the BA scheme may take any value in the set $\mathcal{U}$. The expected value of the average beamwidth resulted from the BA scheme for policy $\mathcal{P}$ is 
\begin{align}
    &\bar{U}_{\mathcal{P}}(\mathcal{B})=\sum_{j=1}^{N} c_{j} \mathbb{E}\left[\left|B_{\mathcal{P}}(\mathbf{s})\right|\right]\label{init_obj}, \quad \text{ where,} \\
    &\mathbb{E}\left[\left|B_{\mathcal{P}}(\mathbf{s})\right|\right]=\sum_{k=1}^{M}\left|u_{k}\right| \mathbb{P}\left\{B_{\mathcal{P}}(\mathbf{s})=u_{k}\right\}
\end{align}
and $|u_k|$ denotes the Lebesgue measure of the $u_k$. Note that $u_k$ may be a finite union of multiple intervals in which case $|u_k|$ will be the sum of their widths. Given the value of $b$ the objective of the BA scheme is to design $\left\{\Phi_{i}\right\}_{i=1}^{b}$ such that the expected average TB beamwidths as in \eqref{init_obj} gets minimized. i.e., 
\begin{align}
    \left\{\Phi_{i}^{*}\right\}_{i=1}^{b}={\arg \min }_{{\left\{\Phi_{i}\right\}_{i=1}^{b}}} \bar{U}_{\mathcal{P}}\left(\left\{\Phi_{i}\right\}_{i=1}^{b}\right)
\end{align}

As shown in \cite{Khal20}, it is straightforward to establish that a multi-user NI-BA problem can be posed as single-user NI-BA by casting the the weighted average of the users' PDFs as a prior on the AoD of a single user. 
\begin{align}
    f_{\Psi}(\psi)=\sum_{j=1}^{N} c_{j} f_{\Psi_{j}}(\psi), \quad \psi \in \mathcal{D}\label{sing-user-pdf}
\end{align}

Therefore, we solve the problem for the single-user case with the PDF as in \eqref{sing-user-pdf} and remove the index $j$ from the notations. 

Let $P_A$ be the probability of receiving a binary feedback sequence with the positivity set $A$. Using the inclusion-exclusion principle we can express $P_A$ as follows, 
\begin{align}
P_{A} &=\left(\sum_{C \subset A} g\left(\omega_{C}\right)\right)^{p}-\sum_{B \subset A^{(L-1)}}\left(\sum_{C \subset B} g\left(\omega_{C}\right)\right)^{p} \nonumber\\
& +\sum_{B \subset A^{(L-2)}}\left(\sum_{C \subset B} g\left(\omega_{C}\right)\right)^{p}-\ldots \nonumber\\
& +(-1)^{(L+1)} \sum_{B \subset A^{(1)}}\left(\sum_{C \subset B} g\left(\omega_{C}\right)\right)^{p}
\end{align}
where $g\left(\omega_{C}\right)=\int_{\psi \in \omega_{C}} f_{\Psi}(\psi) d \psi$, and $A^{(\ell)}, \ell \in [L]$ is the set of all subsets of $A$ with size $\ell$. Further, let $\lambda_{\mathcal{P}}(A)$ be the width of the TB resulted from the feedback sequence $\s$ with the positivity set $A$. The objective function \eqref{init_obj} can be rewritten as, 
\begin{align}
    \Bar{\lambda}\doteq \sum_{A \subset[b]} \lambda_{\mathcal{P}}(A) P_{A}. \label{eq:lambda}
\end{align}
where $\lambda_{\mathcal{P}}(A)$ for mentioned policies is expressed as,
\begin{align}
    &\lambda_{SD}(A)=\sum_{C \subset A} \lambda\left(\omega_{C}\right)\label{l_sd}\\
    &\lambda_{BF}(A)=\min _{i \in A} \sum_{C, i \in C, C \subset A} \lambda\left(\omega_{C}\right)\label{l_bf}\\
    &\lambda_{2-SD}(A) = \sum_ {V \in \mathcal{V}_A} \lambda\left(\omega_{V}\right) \label{l_sd2}\\
    &\lambda_{2-BF}(A) = \sum_ {i=1}^n \lambda\left(\omega_{t_i}\right), \text{ where } T^* = (t_1, \ldots, t_n) \label{l_bf2}
\end{align}
%
The optimized scanning beam set $\mathcal{B}^*$ is obtained from $\mathcal{C}^*$ where
\begin{align}
    \mathcal{C}^* = \arg\min _{\mathcal{C}} \Bar{\lambda} \label{opt_form}
\end{align}

\section{Proposed Beam Alignment Scheme}
\label{sec:proposed}

In this section, we propose our solution to the mentioned optimization problem. A set of SB is called \emph{generalized exhaustive search (GES)} if and only if for any $i$ and $j$, $\Phi_{i} \cap \Phi_{j}=\varnothing$. A set of SB is called \emph{exhaustive search (ES)} if and only if it is GES and $\lambda\left(\omega_{i}\right)=\lambda\left(\omega_{j}\right)$. A contiguous beam is denoted by its angular coverage interval (ACI), e.g., the beam $\Phi_i$ is denoted as $[s_i,e_i)$. A \emph{composite beam} is defined as a beam with multiple disjoint ACIs. As the number of ACIs increases, the sharpness of the beams deteriorates. For the scanning beams it is desirable to use the sharpest beams, hence, we use contiguous beams (beam with single ACIs) as scanning beams. It is not hard to show that $b$ scanning beams generates at most $2b$ CBs due to possible intersection of multiple scanning beams. Out of possible set of scanning beams, some are more appropriate. Since, the policy is a function from the set of feedback sequences, it is desirable to maximize the size of the set of feedback sequences. Hence, we first pose the following question: ``What is \emph{the most distinguishable} set of scanning beams, i.e., the set of beams which can generate the maximum number of possible feedback sequences?"

To answer this question, we define a special form for the set of scanning beams, namely, \emph{Tulip design} for which we have proved it generates the maximum number of feedback sequences for $p=1$ and $p=2$. While, we strongly believe that the same is true for $p \geq 3$, we do not have a formal proof. Hence, any results that is presented in the evaluation section for $p \geq 3$ is merely the results obtained under the assumption of using Tulip design.






\begin{defn}
Tulip design is given by a set of contiguous SBs $\mathcal{B}=\left\{\Phi_{i}\right\}, i \in[b]$ where each beam may only have intersection with its adjacent beams with the exception of $\Phi_1$ and $\Phi_b$ for which the intersection might be nonempty. This means $\Phi_i \cap \Phi_j = \varnothing$, $1 <|i-j|< b-1$.
\end{defn}




\begin{thm}
Among the set of contiguous scanning beams, a set of scanning beams with Tulip design generates the maximal number of possible feedback sequences for the channel with $p = 1$ and $2$, for an arbitrary distribution of channel AoD that is nonzero on any points in the range $[0, 2 \pi)$.
\label{lemma:Tulip_2}
\end{thm}

\proof
Please see \cite{NEC_url}.

Under the Tulip design, the CB set takes a special form given by $\mathcal{C}_{\text{eff}} = \mathcal{C}^1_{\text{eff}} \cup \mathcal{C}^2_{\text{eff}}$ where $\mathcal{C}^1_{\text{eff}} = \{\omega_i\}_{i=1}^b$ and $\mathcal{C}^2_{\text{eff}} = \{\omega_{i, i \oplus 1}\}_{i=1}^b$. Clearly, $|\mathcal{C}_{\text{eff}}| = 2b$. 
By using Tulip design, we can reformulate the optimization problem \eqref{opt_form} in terms of the starting and ending point of the ACI of the SB $\Phi_i$, i.e., $\Phi_{i}=\left[x_{i}, y_{i}\right), i \in[b]$. Hence, we have $\omega_i =[y_{i\ominus1}, x_{i\oplus1})$ and $\omega_{i,i\oplus1} = [x_{i}, y_{i\ominus1})$. We have
%
%
\begin{align}
    \mathcal{C}^*_{\text{eff}} &= \arg\min _{\mathcal{C}_{\text{eff}}} \Bar{\lambda} \label{new_eff_obj}\\
    &x_{i+1}\geq x_{i}, \quad \forall i \in[b-1]\label{start_cons}\\
    &y_{i+1}\geq y_{i}, \quad \forall i \in[b-2]\\
    &x_{i+2}\geq y_{i}\geq x_{i+1}, \quad \forall i \in[b-2]\\
    &x_{1}\leq y_{b-1}\leq 2 \pi+x_{1}\\
    &y_{b}\leq 2 \pi+x_{2}\\
    &2 \pi+x_{1}\leq y_{b}\label{validdity_cons}
\end{align}
where constraints \eqref{start_cons}-\eqref{validdity_cons} ensure the validity of the Tulip design. The optimization problem \eqref{new_eff_obj} is generally nonlinear. For instance, for uniform distribution on the AoD of the user, the objective function \eqref{new_eff_obj} is a polynomial function of the order $(p+1)$ of the beamwidth of the CBs. We propose a greedy algorithm to solve the BA optimization problem that is pseudo-coded as follows in \emph{Algorithm~\ref{greedy-sa}}. The \emph{Greedy-SA} algorithm starts by discretizing  the angular domain $\mathcal{D} = [0, 2\pi]$ to get the ground set $G_N$, the quantized version of the angular range consisting of $N$ points. It then randomly picks $2b$ points from the ground set and forms the initial CB set $\mathcal{C}_{\text{eff}}$. Hence, the initial value of $\Bar{\lambda}$ can be easily computed using \eqref{eq:lambda}. The \emph{Greedy-SA} algorithm makes repeated calls to the \emph{Modify-Sol} routine pseudo-coded in \emph{Algorithm~\ref{modif_sol}} to improve the quality of the CB set $\mathcal{C}_{\text{eff}}$, i.e. to reduce the value of $\Bar{\lambda}$. Each time the \emph{Modify-Sol} routine is called it performs the following sequence of operations. The routine generates the set $perm$ of all random tuples $(p, q, r)$ where $z_p$ and $z_q$ are two of the points in the set $\{z_i\}_{i=1}^{2b}$ and $r$ denotes a direction in the set $\{forward, backward\}$. It then repeatedly picks one such tuple and then slides the window $\{z_i\}_{i=p}^q$ over the ground set $G_N$ in direction $r$ and computes the new value for $\Bar{\lambda}$, namely $\Bar{\lambda}_{new}$. The first time $\Bar{\lambda}_{new}$ goes lower than its old value, the routine records $\Bar{\lambda}_{new}$ and the corresponding $\mathcal{C}_{\text{eff}}$ and calls itself again with these new values. The \emph{Greedy-SA} algorithm terminates when all the points $\{z_i\}_{i=1}^{2b}$ are stable. In other words, when there are no tuples $(p,q, r) \in perm$ which improves the value $\Bar{\lambda}$ from equation \eqref{eq:lambda} by moving the window $\{z_i\}_{i=p}^q$.

\begin{algorithm}[t]
\caption{\textit{Greedy-SA}}
\label{greedy-sa}
 \begin{algorithmic}[1]
 \renewcommand{\algorithmicrequire}{\textbf{Input:}}
 \renewcommand{\algorithmicensure}{\textbf{Output:}}
 \REQUIRE $N, b, \mathcal{P}, f_{\Psi}(\psi), done = \varnothing$
 \\ 
 \STATE $G_N \doteq$ a set of $N$ points in $[0, 2\pi]$   
 \STATE $\{z_i\}_{i=1}^{2b} \doteq$ a random ordered set of points from $G_N$  
 \STATE $\{z_i\}_{i=1}^{2b} \Leftrightarrow \mathcal{C}_{\text{eff}} \doteq \{(z_i,z_{i+1\mod b}), i \in [2b]\}$
 \STATE Compute $\Bar{\lambda}$ from \eqref{eq:lambda} using $f_{\Psi}(\psi)$ and $\mathcal{C}_{\text{eff}}$
  \WHILE {not $done$} 
  \STATE $(done, \Bar{\lambda}, \mathcal{C}_{\text{eff}})$\text{ = }\emph{modify\text{-}sol}($G_N$, $\Bar{\lambda}$, $\mathcal{C}_{\text{eff}}$)
  \ENDWHILE
 \STATE Return $\Bar{\lambda}$, $\mathcal{C}_{\text{eff}}$
 \end{algorithmic}
\end{algorithm}

\begin{algorithm}[t]
\caption{\textit{Modify\text{-}Sol}}
\label{modif_sol}
 \begin{algorithmic}[1]
 \renewcommand{\algorithmicrequire}{\textbf{Input:}}
 \renewcommand{\algorithmicensure}{\textbf{Output:}}
 \REQUIRE $G_N$, $\Bar{\lambda}$, $\mathcal{C}_{\text{eff}}$, $s = True$, $count = 0$
 \\ 
 \STATE $dir = \{forward, backward\}$, $\Bar{\lambda}_{old} = \Bar{\lambda}$
 \STATE $perm =$ Shuffle $\{(p,q,r)| p,q \in [2b], r \in dir, p \leq q\}$
 \REPEAT{
 \STATE Orderly select next tuple $(p,q,r)$ from $perm$
 \STATE Slide $\{z_i\}_{i=p}^q$ in $r \in dir$ direction on points in $G_N$ 
 \STATE Compute $\Bar{\lambda}_{new}$ from \eqref{eq:lambda} using $f_{\Psi}(\psi)$ and $\mathcal{C}_{\text{eff}}$
 \IF{$\Bar{\lambda}_{new}\geq\Bar{\lambda}_{old}$}
 \STATE $count++$ 
 \ENDIF
 }
 \UNTIL{$(count  = 2b^2 + b) \lor{(\Bar{\lambda}_{new}<\Bar{\lambda}_{old})}$}
\IF{$count = 2b^2 + b$}
 \STATE Return $(True, \Bar{\lambda}_{new}, \mathcal{C}_{\text{eff}})$
 \ELSE
 \STATE Return \emph{modify\text{-}sol} ($G_N$, $\Bar{\lambda}_{new}$, $\mathcal{C}_{\text{eff}}$)
 \ENDIF

 \end{algorithmic}
\end{algorithm}


\section{Performance Evaluation}
\label{sec:evaluation}

In this section, we evaluate the performance of our beam alignment scheme in the multi-path environment for mentioned policies, i.e.,  $\mathcal{P} \in \{SD, BF, p-SD, p-BF\}$, by means of numerical simulations. 
We characterize the solutions for different policies. While we only include limited simulation results due to the space constraint, we strive to provide additional insights based on our extensive simulations. In practice the user channels has only a few resolvable paths, hence, we mainly focus of $p=2,3$. 
We consider uniform distribution, the cut-normal distribution, i.e., $\mathcal{N}(\mu = \pi, \sigma = 1)$ that is truncated beyond the range $(0, 2\pi)$.

\begin{figure}[t]
    \centering
            \subfloat[$SD$ Policy, $\Bar{\lambda} = 2.26$  \label{sd_b5}]{
            \includegraphics[width=0.5\linewidth]{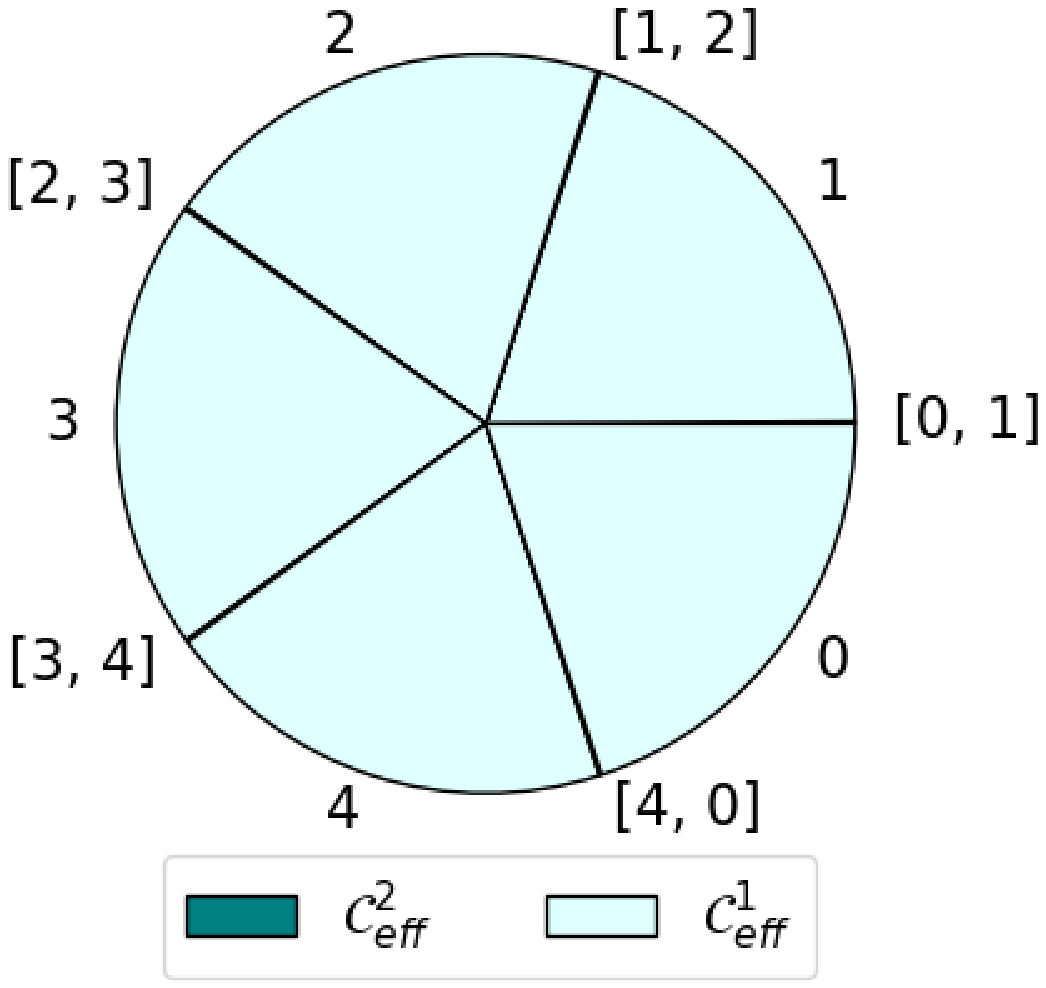}}
            \subfloat[$BF$ Policy, $\Bar{\lambda} = 1.145$ \label{bf_b5}]{
            \includegraphics[width=0.5\linewidth]{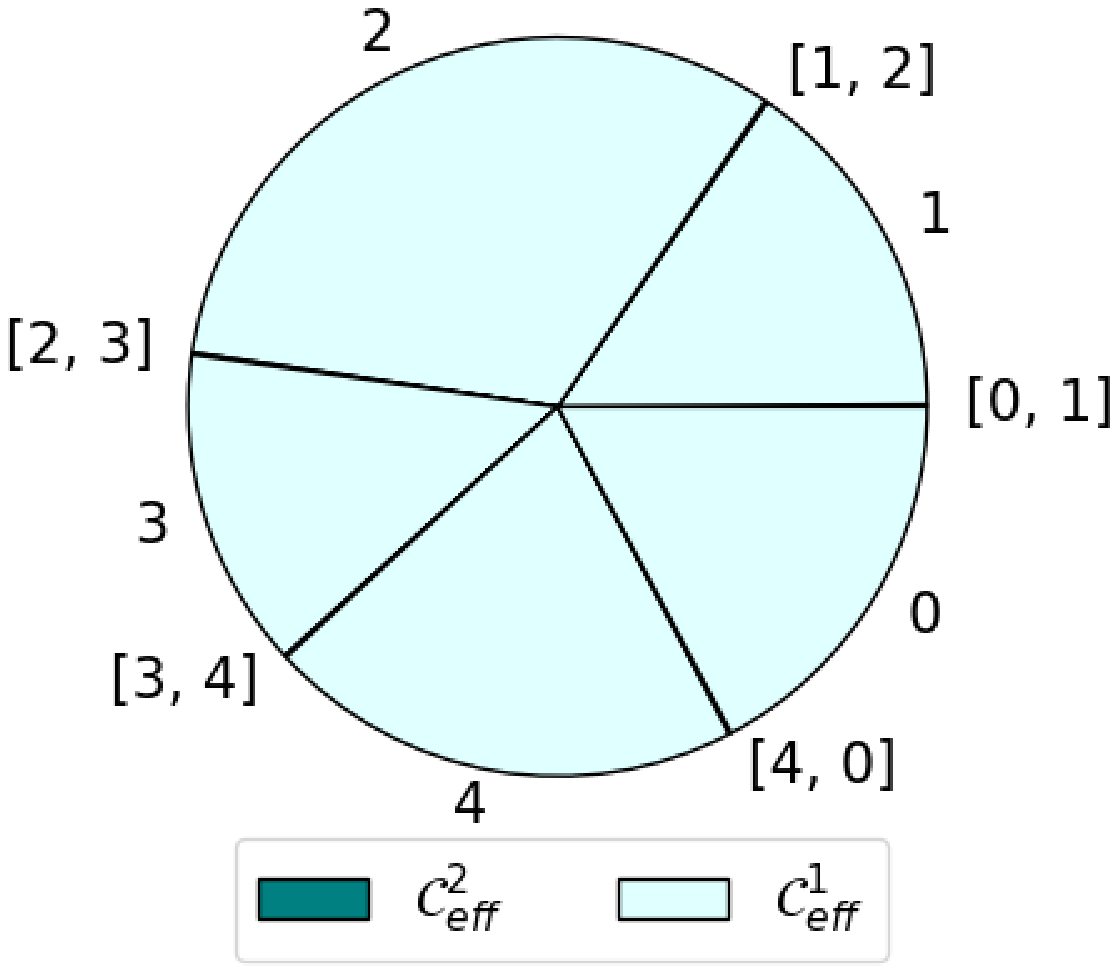}}
            \caption{$p=2$, $b=5$, $N= 1000$, Uniform PDF}
            \label{fig:uniform_p2}
\end{figure}

\begin{figure}[t]
    \centering
            \subfloat[$2-SD$ Policy, $\Bar{\lambda} = 1.822$ \label{sd_2_b5}]{
            \includegraphics[width=0.5\linewidth]{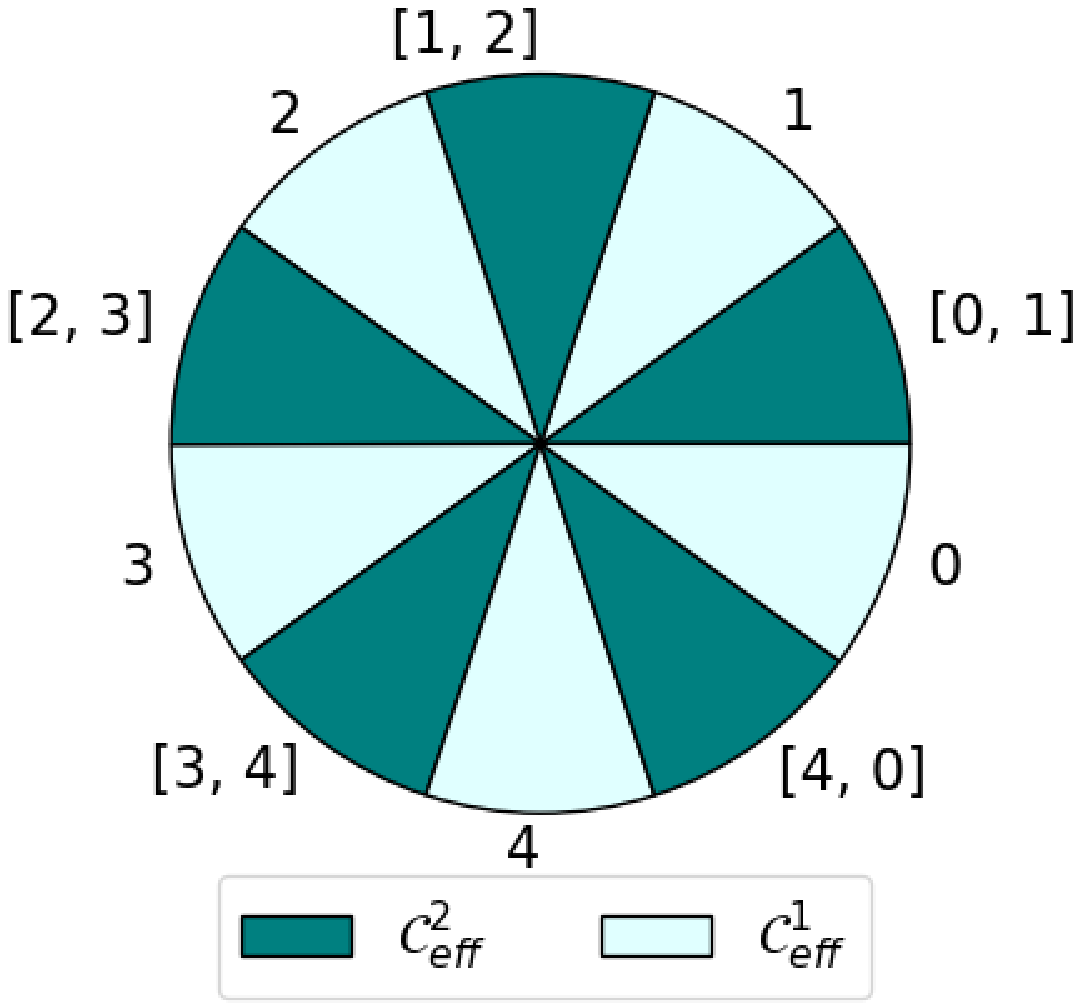}}
            \subfloat[$2-BF$ Policy, $\Bar{\lambda} = 0.836$ \label{bf_2_b5}]{
            \includegraphics[width=0.5\linewidth]{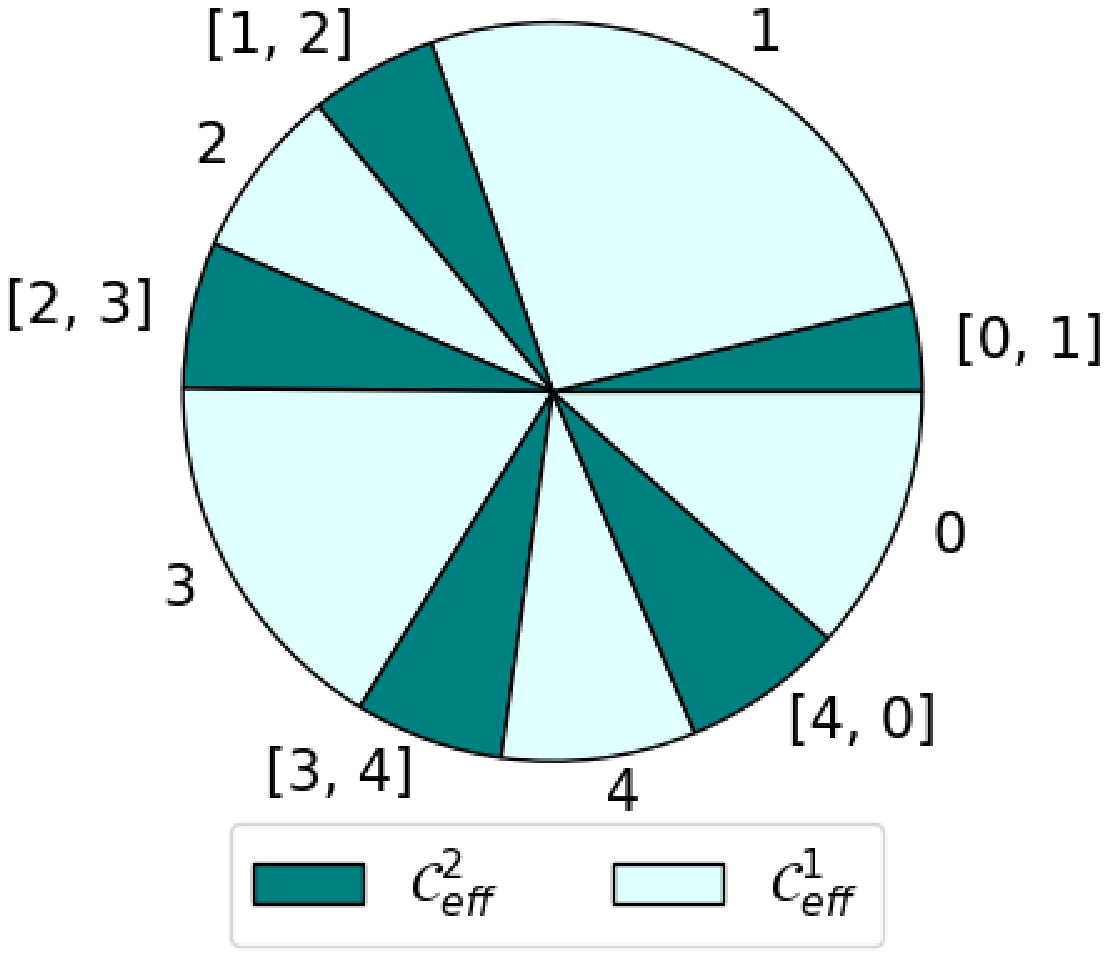}}
            \caption{$p=2$, $b=5$, $N= 1000$, Uniform PDF}
            \label{fig:uniform_p2_2}
\end{figure}

\begin{figure}[t]
    \centering
            \subfloat[$SD$ Policy, $\Bar{\lambda} = 1.76$  \label{sd_b5}]{
            \includegraphics[width=0.5\linewidth]{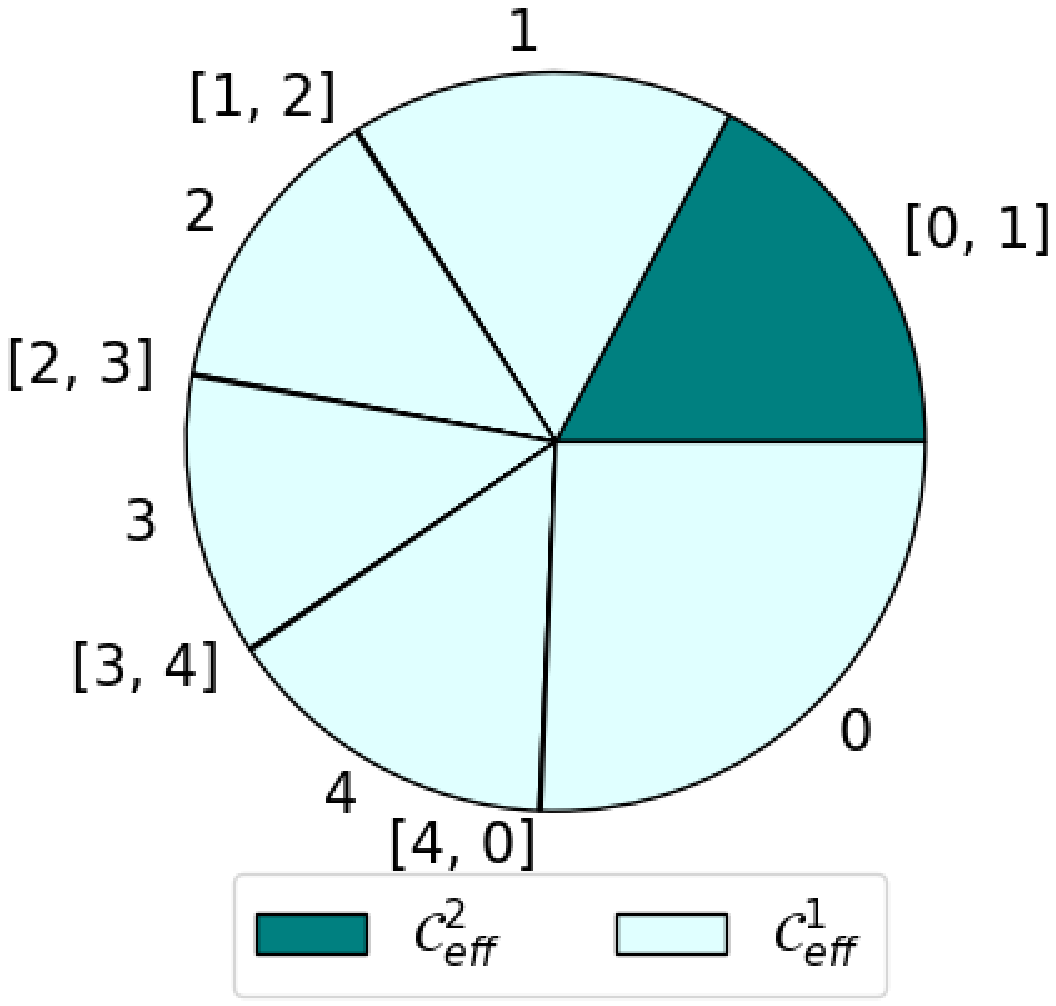}}
            \subfloat[$BF$ Policy, $\Bar{\lambda} = 0.71$ \label{bf_b5}]{
            \includegraphics[width=0.5\linewidth]{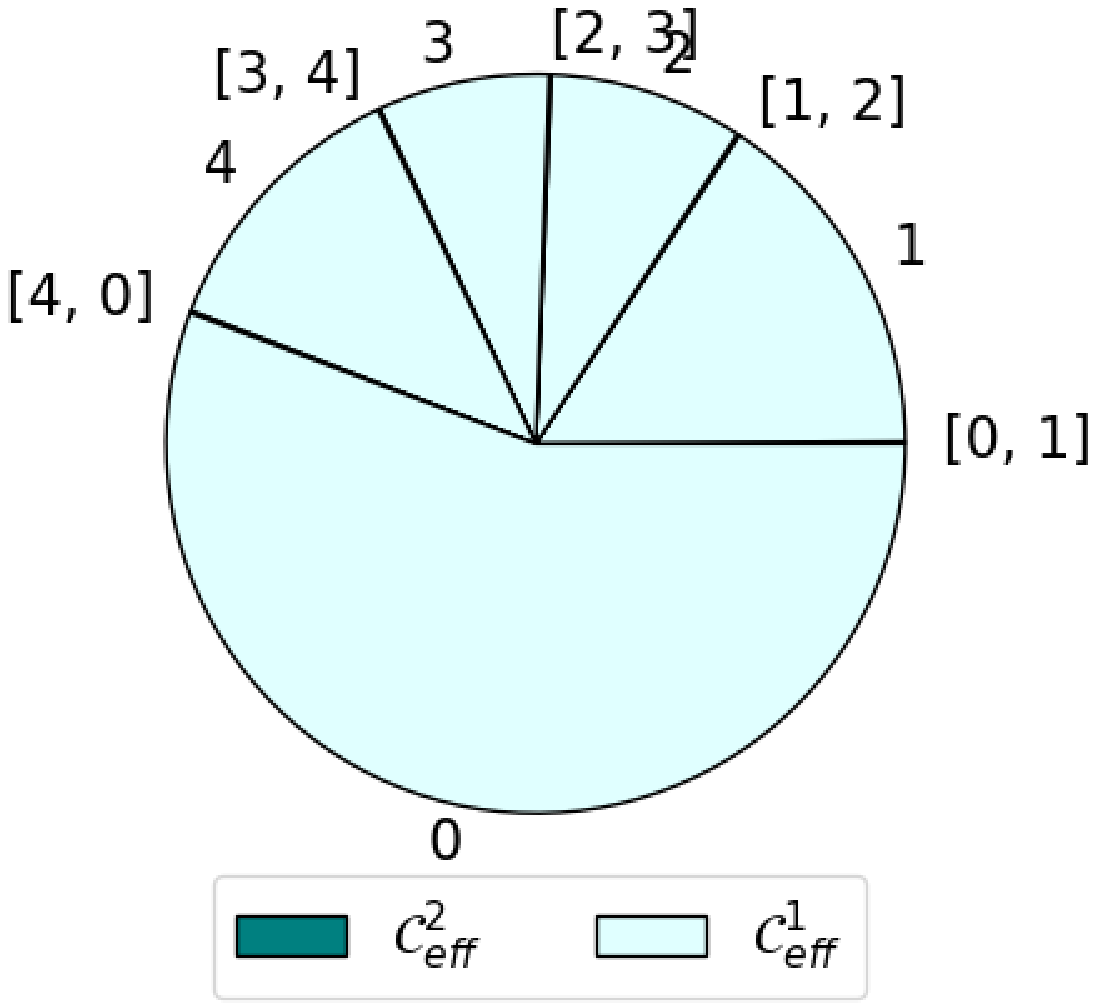}}
            \caption{$p=2$, $b=5$, $N= 1000$, Cut-Normal PDF}
            \label{fig:normal_p2}
\end{figure}

Fig.~\ref{fig:uniform_p2}-\ref{fig:normal_p2} demonstrate the result of BA under the Tulip design for $b=5$ and $p=2$. Each figure shows the CB set that is the output of running algorithm~\ref{greedy-sa} under the given BA policy and the given average PDF of all users. The labels corresponding to each CB are tagged on the corresponding arc. 

Fig.~\ref{fig:uniform_p2} depicts the result of the BA scheme under the $SD$ and the $BF$ policies where uniform angular distribution is considered for the AoD of the users. This PDF can be interpreted as having no prior information on the AoD of the users.  
As intuitively expected, it is observed that the solution to the BA under the $SD$ policy is an ES. In other words, the component beams in the set $\mathcal{C}^2_{\text{eff}}$ take zero lengths and the beams in $\mathcal{C}^1_{\text{eff}}$ take equal lengths.  
We observe that the solution to the BA under the $BF$ policy is not ES anymore but it takes the form of a GES. Through extensive simulations, we have observed that, for uniform distribution and $b>2$, the solution to BF and SD policies are always a GES, and ES, respectively. However, it is not trivial to analytically prove this result. Please note that the solution for $BF$ policy and $b=2$ is not GES. One can easily verify that the optimal solution for BF policy is not unique as any permutations of the $b$ arc would have the same $\Bar{\lambda}$, nonetheless, the minimum value of $\Bar{\lambda}$ is unique. Indeed, the presented greedy algorithm always converges to the same minimum value for any initialization. We note that the average expected TB beamwidth under the $BF$ policy is less than that of the $SD$ policy which means that the $BF$ policy has a higher beamforming gain. Given that the solution to BF is GES and the permutations of the arc is not important, one can write the objective function explicitly in terms of the length of the arcs in the form of a polynomial of degree $b$. The solution which minimizes this polynomial can be numerically found. For example, for BF policy, the solution in terms of the length of the arcs for $b=3,4,5$ is $(1.36,1.93,2.99)$, $(1.13,1.27,1.52,2.36)$, $(0.89,0.97,1.09,1.31,2.02)$ with minimum values of $1.94$, $1.44$, $1.14$, respectively. While in SD policy the length of the arcs are equal, in BF policy the length of the arcs vary due to the fact that in BF policy we take the arc with minimum ACI when we receive positive feedback on more than one SB. 

Fig.~\ref{fig:uniform_p2_2} depicts the BA result under the $p\text{-}SD$ and the $p\text{-}BF$ policies for $p=2$. It is observed that under the $2\text{-}SD$ policy all the CBs take equal length. It should be pointed out that for the $2\text{-}BF$ policy the greedy algorithm~\ref{greedy-sa} converges to multiple (but usually a small number of) solutions by using different initialization. Hence, we take the minimum of such solution. We observe that the value of $\Bar{\lambda}$ decreases under the $2\text{-}SD$ and the $2\text{-}BF$ policies comparing to that of the $SD$ and the $BF$ policies and this is due to the luxury of having the information on the value of $p$.

Fig.~\ref{fig:normal_p2} shows the result of BA when a cut-normal PDF $\mathcal{N}(\pi,1)$ truncated between  $[0,2\pi)$ is considered. 
It is observed that the $SD$ policy is not GES anymore while the $BF$ policy still generates a GES solution. 
A cut normal distribution on average PDF of the users may be interpreted as having prior knowledge about the users AoDs. For example the mean and variance of the users AOD from prior beam alignment frames may be used to fit a cut normal distribution on the PDF of the users' AoDs in the current frame. Comparing $\Bar{\lambda}$ for uniform and cut-normal distributions both for $SD$ and $BF$ policies illustrates that the prior knowledge about the users' AoDs results in a smaller $\Bar{\lambda}$.

\begin{figure}
\begin{center}
\begin{minipage}[t]{0.48\textwidth}
\includegraphics[width=1\textwidth]{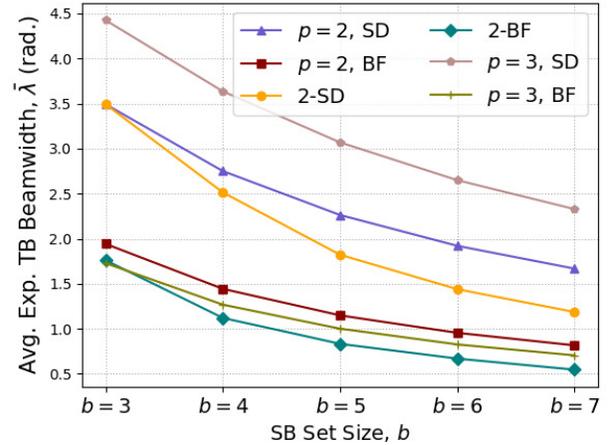}
  \caption{Impact of varying the size of SB set on the BA solution}
  \label{fig:ba_scheme}
\end{minipage}
\end{center}
\end{figure}

Fig.~\ref{fig:ba_scheme} demonstrates the expected average beamwidth of the users, $\Bar{\lambda}$, for different policies as a function of the size of the SB set $b$. For the channels with uniform distribution and $p=2$ paths, it is observed that $BF$ policy results in a smaller $\Bar{\lambda}$ than the $SD$ policy, hence, achieves higher beamforming gain. The same is true for the channel with $p=3$. Notably, if the channel has larger number of paths, say $p=3$ vs. $p=2$, the $\Bar{\lambda}$ for $BF$ policy decreases as having additional paths allows for choosing the minimum among additional possibilities. However, the $SD$ policy would need to cover more paths and hence has a larger $\Bar{\lambda}$. Fig.~\ref{fig:ba_scheme} also shows that using the information about the number of paths in $2-SD$ and $2-BF$ policies versus $SD$ and $BF$ policies would result in a better performance, i.e., smaller $\Bar{\lambda}$, respectively. Finally, we note that $\Bar{\lambda}$ reduces for all policies as the number of scanning beam increases.



\section{Conclusions}
\label{sec:conclusions}
We studied the non-interactive multi-user beam alignment problem in mmWave systems while considering the effect of multi-path. We introduced the \emph{Tulip design} for the scanning beams in the probing phase and proved its optimality in terms of maximizing the achievable number of feedback sequences when the scanning beams are contiguous. We modeled beam alignment as an optimization problem under different policies and proposed a greedy algorithm to find the optimal BA scheme. We characterized the solutions of our BA scheme by means of numerical experiments and presented our observations.  


\renewcommand{\nariman}[1]{\textcolor{red}{#1}}
\renewcommand{\amir}[1]{\textcolor{blue}{#1}}



%

\bibliographystyle{IEEEtran}
\bibliography{bibliography}

\end{document}